\documentclass[
aps,%
12pt,%
final,%
notitlepage,%
oneside,%
onecolumn,%
nobibnotes,%
nofootinbib,%
superscriptaddress,%
noshowpacs,%
centertags,%
amsmath,amssymb]%
{revtex4}
\usepackage{graphicx}

\begin{document}
\title{Elastic $Nd$ scattering at intermediate
energies as a tool for probing the short-range deuteron structure}
\author{\firstname{M.~A.}~\surname{Shikhalev}}
\email{shikhalev@jinr.ru} \affiliation{Joint Institute for Nuclear Research, 141980 Dubna,
Russia}

\begin{abstract}
A calculation of the deuteron polarization observables $A^d_y$, $A_{yy}$, $A_{xx}$, $A_{xz}$ and
the differential cross-section for elastic nucleon-deuteron scattering at incident deuteron
energies $270$ and $880$ MeV in lab is presented. A comparison of the calculations with two
different deuteron wave-functions derived from the Bonn-CD $NN$-potential model and the dressed
bag quark model is carried out. A model-independent approach, based on an optical potential
framework, is used in which a nucleon-nucleon $T$-matrix is assumed to be local and taken on the
energy shell, but still depends on the internal nucleon momentum in a deuteron.
\end{abstract}

\maketitle

\section{Introduction}
The reaction of elastic nucleon-deuteron scattering is considered both by experimentalists and
theoreticians as one of the clue tasks in few-nucleon physics. For three decades it has been
served as a hope to obtain more information about the intermediate- and short-range $NN$
interaction and as a probe of the deuteron structure at small distances (for a review, see Ref.
\cite{Gloekle}). During the last decade it is also studied with the purpose of testing of
various three-nucleon forces (3NF) and particularly their spin-dependence
\cite{Sekiguchi,Uzikov}. It is also of interest as a basic reaction to establish a polarimetry
for vector-tensor mixed polarized beams.

To describe $Nd$ elastic scattering below the pion production threshold different techniques
have been applied \cite{Friar,Kievsky,Witala,Mach}. The momentum space Faddeev equations can now
be solved with high accuracy for the most modern two- and three-nucleon forces. The 3NFs in such
a calculations are of Fujita-Miyazawa \cite{Fujita,Urbana} or Tucson-Melbourne \cite{Coon} type.
It was found from these calculations that the differential cross section and the polarization
observables are essentially insensitive to the choice of the two-nucleon interaction provided it
is in agreement with $NN$ elastic scattering data.

The next step is to go to the higher momentum transfer region and to explore this reaction above
the $NN$ inelastic threshold. But calculations at incident proton energies greater than $200$
MeV in lab encounter with some nontrivial difficulties. The first problem is that up to now
there is no reliable quantitative model for the $NN$-interaction above the inelastic threshold.
All existing models that pretend to description of two-nucleon scattering up to 1 GeV do it only
in semi-quantitative  manner \cite{Mach2,Elster}. The second problem is that it is no longer
conventional three-body problem. Nucleon isobar and meson degrees of freedom start to play a
significant role and disguise the effects from the short-range two-nucleon interaction
\cite{Uzikov2}. And last but not least, there are purely computational difficulties in
performing relativistic Faddeev calculations at higher energies \cite{Gloeckle2,Gloeckle3}.
Hence, the existing theoretical frameworks of elastic $Nd$ scattering at intermediate energies
concern mainly to the very backward scattering angles where short-range behavior of a deuteron
wave function (DWF) is, as expected, the most evident. The common framework at these energies is
a multiple scattering formalism, where the two-body $Nd$ amplitude is expanded in series of a
$NN$ T-matrix. In the absence of a reliable two-nucleon potential in the GeV region, the
T-matrix is usually taken either on-energy-shell \cite{Alberi} or extrapolated off-shell in a
model independent fashion \cite{Ladygina}. So, the DWF contains the only information about the
$NN$ interaction in these models.

Now, there is a set of high quality nucleon-nucleon potentials, based mainly on a meson exchange
picture, that describe deuteron properties and two-nucleon elastic scattering data perfectly
below the pion production threshold \cite{Stocks,Wiringa,Machleidt}. With the inclusion of a
3NF, they also provide a good description for the 3N binding energies and the nucleon-deuteron
differential cross section. The general feature of these models is an uniform depletion of the
DWF at small internucleon distances.

In the last two decades, numerous attempts have been made to describe the intermediate and
short-range $NN$ interaction by methods of QCD \cite{st,VM1,Yaz,Sal,Kuk}. The advantages of the
quark models are that only a few physically meaningful parameters can be used to describe
processes involving hadrons and a possibility to take the underlying symmetries of QCD directly
into account. Most of these attempts focused only on elastic $NN$ scattering at rather low
energies and/or on low partial waves and are not suitable as input in few-body calculations.
However, some of the features of a few-nucleon system, derived within the frameworks, can be
further explored and tested. Especially it concerns the short-range behavior of the deuteron and
3N bound states. One of the successful models based on six-quark ($6q$) symmetries is a dressed
bag model (DBM) proposed recently \cite{Kuk,Kuk2}. The essence of this model lies in different
dynamics of the quark configurations $|s^4p^2[42]_xL=0,2\rangle$, $|s^3p^3[33]_xL=1,3\rangle$
and the most symmetric ones $|s^6[6]_xL=0\rangle$, $|s^5p^1[51]_xL=1\rangle$. Whereas the first
two configurations have a cluster structure and are projected mainly on the $NN$ channel, the
third and fourth configurations have the structure of a quark bag with a large weight of the
$\Delta\Delta$ and $CC$ (hidden-color) states. As a result, the deuteron and 3N wave functions
have a short-range node, which result from the orthogonality of the cluster wave function and
the wave function of the  $6q$ compound state. These $6q$ symmetry arguments are also the
foundations for the Moscow $NN$-potential model \cite{Kuk3,Knyr}. In the framework of the DBM,
it turned out to be possible to fit very reasonably the $NN$ phase shifts in $^1S_0$ and
$^3S_1-^3D_1$ channels up to 1 GeV and the deuteron static properties as well. As to the 3N
systems, this model could explain quantitatively all static properties of $^3\rm{He}$ and
$^3\rm{H}$ ground states, including a precise parameter-free description of the Coulomb
displacement energy of $^3\rm{He}-{}^3\rm{H}$ and all the charge distributions in these nuclei
\cite{Kuk4,Kuk5}.

In this work, elastic $Nd$ scattering is considered as a possible discriminative tool between
calculations with two different kinds of a DWF. The first one is derived from the Bonn-CD $NN$
potential \cite{Machleidt} and diminishes uniformly with approaching the internucleon distance
to zero. The second one is a result of the DBM \cite{Kuk} and develops a nodal behavior. The
framework is based on an optical potential formalism. A multiple scattering expansion is used to
derive the optical potential that comprises of the one-nucleon-exchange (ONE) mechanism, the
single- and double-scattering terms. As shown by Faddeev calculations, at energies below about
200 MeV rescattering of higher order is very important, however around 300 MeV the first two
terms in the expansion are sufficient to describe the total $Nd$ cross section \cite{Witala2}.
For the $NN$ input, a model-independent approach is used in which the nucleon-nucleon $T$-matrix
is assumed to be local and taken on-shell, but in contrast to the common impulse approximation
it still depends on the internal nucleon momentum. Thus the two-nucleon amplitude cannot be
factorized out of an integral and some kinematical off-shell effects are implicitly taken into
account.

The deuteron vector and tensor analyzing powers and the differential cross section are
calculated for two deuteron kinetic energies $E_d$ in lab - 270 MeV and 880 MeV. At $E_d=270$
MeV a comparison with precise data measured at RIKEN \cite{RIKEN} is performed to validate the
model. At this energy the difference between the calculations with the two kinds of a DWF is, as
expected, not remarkable. The calculation at 880 Mev is performed in view of the recent
experiment at JINR \cite{JINR}, in which the deuteron polarization observables were measured in
the region of the so called "cross-section minimum" ($\theta_{\rm{cm}}=70-140^\circ$) where both
vector and tensor analyzing powers can have large values.

The structure of the paper is as follows. In Section 2, the formalism of the optical potential
and multiple series expansion is briefly given. The approximation to the fully off-shell $NN$
T-matrix is developed and the $Nd$ scattering amplitude is defined. Section 3 is devoted to
results of the calculation and their discussion. Conclusion summarizes the content of the work.

\section{Theoretical framework}
The basic equations for a three-body system in quantum physics are Faddeev equations which can
be written in the operator form \cite{Gloekle}:
\begin{equation}
\label{Fadd}
 U=PG_0^{-1}+PTG_0U,
\end{equation}
where $U=U_{\mu_d'\mu_N',\mu_d\mu_N}(\vec{q}\,',\vec{q})$ is an amplitude of elastic
$Nd$-scattering, $\vec{q}$ ($\vec{q}\,'$) -- initial (final) relative momentum in the
nucleon-deuteron c.m., $\mu_d,\mu_N$ are spin quantum numbers, $T$ -- $NN$ scattering T-matrix,
$G_0=(E-H_0+i\epsilon)^{-1}$ is a free propagator of the $3N$ system and $P\equiv
P_{12}P_{23}+P_{13}P_{23}$ stands for a permutation operator that takes into account the
property of identity of three nucleons.

One can rewrite (\ref{Fadd}) in the form that is more appropriate for calculations at higher
energies where a multiple scattering expansion is justified \cite{Kuros}:
\begin{equation}
\label{Optic} U=V_{\rm opt}+V_{\rm opt}G_dU.
\end{equation}
This is an optical potential framework and $V_{\rm opt}$ is a nucleon-deuteron optical
potential:
\begin{equation}
\label{potent}
 V_{\rm opt}=PG_0^{-1}+PT_cG_0V_{\rm
opt}.
\end{equation}
Here $T_c$ is a $NN$ T-matrix without the deuteron pole term and $G_d$ is a deuteron
contribution in the spectral decomposition of the two-body Hamiltonian.

\begin{figure}[t]
\begin{center}
\includegraphics[width=15cm,keepaspectratio]{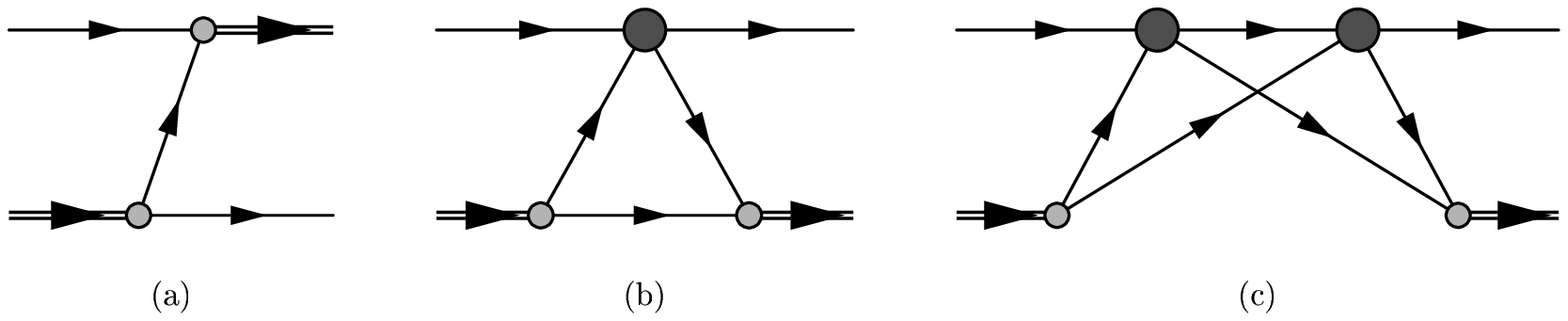}
\caption{The $Nd$ optical potential up to the second order in the multiple series expansion:\\
(a) -- one-nucleon exchange, (b) -- single scattering, (c) -- double scattering; dark-filled
circles represent the $NN$ T-matrix and grey-filled -- deuteron wave function.\label{fig1}}
\end{center}
\end{figure}

Although, Eq. (\ref{Optic}) seems quite simple and represents a common two-body scattering
equation, the derivation of the optical potential from Eq. (\ref{potent}) is as difficult as
solving the Faddeev equations themselves. However, one can employ a multiple scattering
expansion of $V_{\rm opt}$, keeping a possibility to include in $V_{\rm opt}$ some additional
terms originated, for example, due to a 3NF. At higher energies only a few terms in this
expansion may be sufficient to describe observables \cite{Witala2}:
\begin{equation}
\label{Expan}
V_{\rm opt}=PG_0^{-1}+PT_cP+PT_cG_0PT_cP+\ldots
\end{equation}
In Fig. \ref{fig1} a schematic representation for the optical potential (up to the second order
in series expansion) is shown. The first diagram (Fig. \ref{fig1}(a)) is a usual one-nucleon
exchange mechanism, the second graph (Fig. \ref{fig1}(b)) is a so called triangle diagram and
represents the single-scattering process. And Fig. \ref{fig1}(c) represents the
double-scattering which takes into account the break up channel and the double-charge exchange
mechanism.

If to define an initial state as $|i\rangle\equiv|\vec{q};\mu_d\mu_N\rangle_{1(23)}$, where a
proton 1 is in continuum and a proton 2 and a neutron 3 are bound in the deuteron, then
(\ref{Expan}) can be written in the matrix form:
\begin{multline}
\langle f|V_{\rm opt}|i\rangle={}_{2(31)}\langle f|G_0^{-1}|i\rangle+{}_{1(23)}\langle
f|T_c^{2(31)}|i\rangle+{}_{1(23)}\langle f|\widetilde{T_c}|i\rangle+\\
{}_{1(23)}\langle f|\widetilde{T_c}G_0T_c^{2(31)}|i\rangle+{}_{1(23)}\langle
f|T_c^{2(31)}G_0\widetilde{T_c}|i\rangle+{}_{2(31)}\langle f|T_c^{1(23)}G_0T_c^{2(31)}|i\rangle.
\end{multline}
Here,
$${}_{1(23)}\langle f|\widetilde{T_c}\equiv{}_{1(23)}\langle f|T_c^{3(12)}+{}_{2(31)}\langle
f|T_c^{3(12)}$$ is an antisymmetrized proton-proton T-matrix and $T_c^{k(ij)}$ means that the
interaction is occurred between the particles $i$ and $j$ whereas the particle $k$ is a
spectator.

The one-nucleon exchange contribution is written in a usual way:
\begin{equation*}
{}_{2(31)}\langle
f|G_0^{-1}|i\rangle=\left[\sqrt{q^2+M_N^2}+\sqrt{4q^2\cos^2\frac{\theta}{2}+M_N^2}-\sqrt{q^2+M_d^2}
\right]\Psi_{13}^\dag\left(\vec{q}+\frac{1}{2}\vec{q}\,'\right)\Psi_{23}
\left(\vec{q}\,'+\frac{1}{2}\vec{q}\right).
\end{equation*}
Here, a complicated notation related to a summation over spin and orbital quantum numbers is
omitted. $\Psi_{ij}$ is a wave function of the deuteron composed of nucleons $i$ and $j$,
$\theta$ is a scattering angle and
$$q'=q=\left(\frac{E_dM_N^2(E_d+2M_d)}{(M_N+M_d)^2+2M_NE_d}\right)^{1/2},$$
where $E_d$ is a kinetic energy of the deuteron in lab.

To evaluate a single scattering diagram, one must implement an integration over the internal
momentum of nucleons in the deuteron. To do this, a knowledge about the fully off-shell behavior
of the $NN$ T-matrix is required:
\begin{equation}
\label{Sing_scatt} {}_{1(23)}\langle f|T_c|i\rangle=\int\frac{\rm{d}^3p}{(2\pi)^3}
\Psi_{23}^\dag\left(\vec{p}+\frac{1}{4}\vec{k}\right)
T_c\left(\vec{q}\,',\vec{p}-\frac{3}{4}\vec{q}\,'+ \frac{1}{4}\vec{q};
\vec{q},\vec{p}-\frac{3}{4}\vec{q}+ \frac{1}{4}\vec{q}\,'\right)
\Psi_{23}\left(\vec{p}-\frac{1}{4}\vec{k}\right),
\end{equation}
here $k=\vec{q}-\vec{q}\,'$ is a transferred momentum.

However, the absence of a high quality $NN$ interaction model above the inelastic threshold at
the present moment forces anyone to make some approximate evaluation of the integral, proceeding
from the assumption of either off-shell behavior of somehow parameterized T-matrix or taking
only its on-shell value. The common approach of twenty years old calculations is an optimal
impulse approximation \cite{Gurvitz,Neil} in which the internal momentum in the T-matrix is put
to zero in (\ref{Sing_scatt}) that permits the T-matrix to be shifted outside the integral and
the remaining integration produces the deuteron form factor. In this approximation the leading
order corrections due to Fermi motion is vanished provided that the $NN$ amplitude is spin
independent and local. Thus the variation of the $NN$ T-matrix with momentum $\vec{p}$ is
compensated to leading order in $\vec{p}/M_N$ by appropriate choice of the energy parameter upon
which the T-matrix depends. The condition on the energy is that the $NN$ T-matrix be on-shell
when evaluated at $\vec{p}=0$ in Eq. (\ref{Sing_scatt}). However, this  approximation misses
some vital momentum dependency in the T-matrix and is suitable only for scattering on a heavy
nucleus where the recoil is not significant.

In this work, the T-matrix is put on-shell in a way that minimizes the off-shell corrections.
Firstly, the $NN$ T-matrix should be transferred to the two-nucleon c.m. frame, where it is
usually defined. By means of the Lorenz transformation ${\cal
L}^\mu_\nu(\vec{P}_{\mathrm{cm}})$, one has:
\begin{equation}
\label{transform}
T_c\left(\vec{q}\,',\vec{p}-\frac{3}{4}\vec{q}\,'+ \frac{1}{4}\vec{q};
\vec{q},\vec{p}-\frac{3}{4}\vec{q}+ \frac{1}{4}\vec{q}\,'\right)=\Lambda({\cal L})
T_c^{\mathrm{cm}}\left(\vec{Q}\,',\vec{Q}\right)\Lambda^{-1}({\cal L}),
\end{equation}
where $\vec{Q}\,'=\left({\cal L }^{-1}[\vec{q}\,']-{\cal L
}^{-1}\left[\vec{p}-\frac{3}{4}\vec{q}\,'+ \frac{1}{4}\vec{q}\right]\right)/2$ and
$\vec{Q}=\left({\cal L }^{-1}[\vec{q}]-{\cal L }^{-1}\left[\vec{p}-\frac{3}{4}\vec{q}+
\frac{1}{4}\vec{q}\,'\right]\right)/2$ are final and initial two-nucleon relative momenta in
$NN$ c.m. and $\Lambda$ is a transition operator that includes the boost and the Wigner spin
rotations of the T-matrix; $\vec{P}_{\mathrm{cm}}=\vec{p}+\frac{1}{4}\vec{q}\,'+
\frac{1}{4}\vec{q}$ -- c.m. momentum of the two-nucleon system.

If the T-matrix is not strongly energy-dependent and essentially local, then it depends on two
variables - transferred momentum $(\vec{Q}\,'-\vec{Q})^2$ for the direct $NN$-interaction and
$(\vec{Q}\,'+\vec{Q})^2$ in case of the exchange mechanism. Then one can introduce two new
variables $\widetilde{Q}$ and $\widetilde{\theta}$ that correspond to the on-shell relative
momentum and scattering angle and define them as follows:
\begin{equation}
|\vec{Q}\,'-\vec{Q}|=2\widetilde{Q}\sin(\widetilde{\theta}/2),\ \ \ \
|\vec{Q}\,'+\vec{Q}|=2\widetilde{Q}\cos(\widetilde{\theta}/2).
\end{equation}
Thus, the T-matrix $T_c^{\mathrm{cm}}$ is taken to be on-mass shell and calculated at the
effective two-nucleon energy in lab
\begin{equation}
\label{E}
E_{eff}=\frac{2\widetilde{Q}^2}{M_N}=\frac{\vec{Q}\,'^2+\vec{Q}^2}{M_N}.
\end{equation}
Thus, although the $NN$ T-matrix is evaluated on-shell, it still contains some off-shell
information, since the effective energy $E_{eff}$ and the scattering angle $\widetilde{\theta}$
depends on the off-shell momenta $\vec{Q}\,'$ and $\vec{Q}$. Particularly, the effective energy
depends on the internal nucleon momentum $\vec{p}$, and therefore the T-matrix cannot be driven
outside the integral in Eq. (\ref{Sing_scatt}). Moreover, some off-shell dependency is present
on a relativistic level in the operator $\Lambda$ when the transformation of the T-matrix from
the $NN$ c.m. frame to the nucleon-deuteron c.m. frame is performed according to Eq.
(\ref{transform}). In the relativistic case, this approximation corresponds to the situation
when the T-matrix depends only on the two kinematical invariants -- $t$ and $u$, and its
evaluation is performed via imposing the on-shell condition on the squared total energy
$s=4M_N^2-u-t$.

For the double scattering term one should evaluate a six-dimensional integral over the internal
momentum in the deuteron and the intermediate momentum of the scattered nucleon:
\begin{multline}
\label{double_scatt} {}_{1(23)}\langle
f|T_2G_0T_1|i\rangle=\\
\int\frac{\rm{d}^3p}{(2\pi)^3}\frac{\rm{d}^3q''}{(2\pi)^3}
\Psi_{23}^\dag\left(\vec{p}+\frac{\vec{q}\,''}{2}\right)
T_2\left(\vec{q}\,',\vec{p}+\frac{1}{2}(\vec{q}\,''-\vec{q}\,');
\vec{q}\,''+\frac{1}{2}(\vec{q}+\vec{q}\,'),\vec{p}-\frac{1}{2}(\vec{q}+\vec{q}\,'')\right)G_0\\
\times
T_1\left(\vec{q}\,''+\frac{1}{2}(\vec{q}+\vec{q}\,'),-\vec{p}-\frac{1}{2}(\vec{q}\,'+\vec{q}\,'');
\vec{q},-\vec{p}-\frac{1}{2}(\vec{q}-\vec{q}\,'')\right)
\Psi_{23}\left(\vec{p}-\frac{\vec{q}\,''}{2}\right),
\end{multline}
The $NN$ T-matrices in (\ref{double_scatt}) are approximated in the same manner as for
single-scattering according to Eqs. (\ref{transform})-(\ref{E}). The Green function $G_0$ is
taken in a nonrelativistic form to simplify the evaluation of the pole part of the integral:
\begin{equation}
G_0^{-1}=\frac{q^2}{2M_N}+\frac{q^2}{2M_d}-\frac{(\vec{q}\,''+(\vec{q}+\vec{q}\,')/2)^2}{2M_N}-
\frac{(\vec{p}-(\vec{q}+\vec{q}\,'')/2)^2}{2M_N}-\frac{(\vec{p}+(\vec{q}\,'+\vec{q}\,'')/2)^2}{2M_N}.
\end{equation}
Assuming here $M_d=2M_N$, one can rewrite it as follows
\begin{equation}
G_0^{-1}=\frac{(q^2-\vec{q}\vec{q}\,')/2-R^2}{2M_N},
\end{equation}
where
\begin{equation}
\label{R}
R^2\equiv\frac{3}{2}q''^2+2p^2+\frac{3}{2}\vec{q}\,''(\vec{q}+\vec{q}\,')+\vec{p}(\vec{q}\,'-\vec{q}).
\end{equation}
Then, redefine the variables of integration and instead of the variables $|\vec{q}\,''|$ and
$|\vec{p}|$ introduce two new variables $R$ and $\alpha$ as
\begin{align}
|\vec{q}\,''|=& \sqrt{\frac{2}{3}}R\cosh\beta\cos\alpha,\\
|\vec{p}|=& \sqrt{\frac{1}{2}}R\cosh\beta\sin\alpha.
\end{align}
The expression for $\beta$ can be found by substituting these definitions in Eq. (\ref{R}).
Thus, the evaluation of the pole part of the integral (\ref{double_scatt}) is straightforward.
Here, the both the pole and the principal parts of integration in the double-scattering term is
taken into account. The evaluation of the integrals is performed by means of the Monte-Carlo
simulations, dividing the integration domain on several parts to minimize numerical errors.

The scattering equation (\ref{Optic}), which is here a relativistic Lippmann-Schwinger equation,
is solved in helicity basis employing a $K$-matrix approximation, i.e. only the pole part of the
two-body propagator $G_d$ is remained thus all terms in the equation contain only on-shell
information about the optical potential and the scattering amplitude:
\begin{multline}
\langle\lambda'_d\lambda'_N|U^J(q',q)|\lambda_d\lambda_N\rangle=
\langle\lambda'_d\lambda'_N|V_{\mathrm{opt}}^J(q',q)|\lambda_d\lambda_N\rangle-
i\frac{A(q,q)}{q}\langle\lambda'_d\lambda'_N|V_{\mathrm{opt}}^J(q',q)U^J(q,q)|\lambda_d\lambda_N\rangle+\\
\frac{2}{\pi}{\cal
P}\int\frac{\rm{d}q''}{q^2-q''^2}\biggl(\langle\lambda'_d\lambda'_N|V_{\mathrm{opt}}^J(q',q'')U^J(q'',q)|\lambda_d\lambda_N\rangle
A(q'',q) -
\langle\lambda'_d\lambda'_N|V_{\mathrm{opt}}^J(q',q)U^J(q,q)|\lambda_d\lambda_N\rangle
A(q,q)\biggr),
\end{multline}
where the kinematical factor is
$$A(q'',q)=q''^2\frac{(E_q+E_{q''})((E_q^2+E_{q''}^2)/2-q^2-q''^2-M_N^2-M_d^2)}{E_q^2+E_{q''}^2},$$
and $E_q=\sqrt{q^2+M_N^2}+\sqrt{q^2+M_d^2}$ is a total energy.

Then for the $Nd$ scattering amplitude in spin space one has (the incident particle is going
along the $z$-axis):
\begin{multline}
\label{sum}
\langle\mu'_d\mu'_N|U(q,\theta)|\mu'_d\mu'_N\rangle=4\pi\sum_J\sum_{\lambda'_d,\lambda'_N}(-1)^{\lambda'_d-\mu_d}
(2J+1)d^{1/2\ast}_{\mu'_N,\lambda'_N}(\theta)d^{1\ast}_{\mu'_d,-\lambda'_d}(\theta)\\
\times
d^{J}_{\lambda'_N-\lambda'_d,\mu_N+\mu_d}(\theta)\langle\lambda'_d\lambda'_N|U^J(q,q)|-\mu_d\mu_N\rangle,
\end{multline}
where $J$ is a total angular momentum.
\section{Results and discussion}
The calculation of the deuteron polarization observables $A_y^d,A_{xx},A_{yy},A_{xz}$ and the
differential cross section is performed with two different deuteron wave functions. The first
one is a wave function derived in the meson-exchange Bonn-CD model \cite{Machleidt}. The general
trait of wave functions of this kind, derived from the most modern $NN$ potentials, is their
uniform behavior at small distances. The other possible choice is a wave function with a nodal
behavior. This node corresponds to the so-called forbidden state in a $NN$ system as a
consequence of the six-quark dynamics and the fact that the mostly symmetric six-quark state
$|s^6\rangle$ has a small $NN$ component \cite{Obukh}. As one of the representatives of such a
wave functions with nodal behavior can serve a wave function of the DBM \cite{Kuk}, which has a
node in the $^3S_1$ wave at a distance of $\simeq 0.6$ fm.

\begin{figure}[t]
\begin{center}
\includegraphics[width=17cm,keepaspectratio]{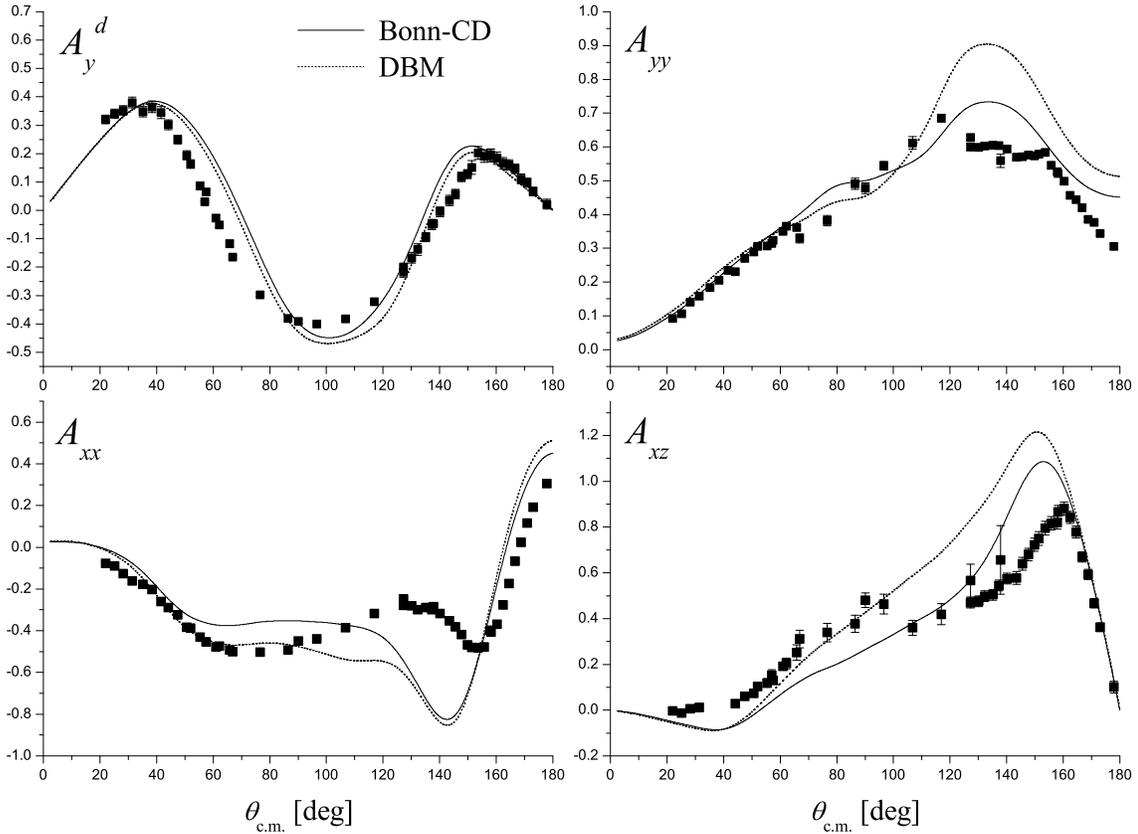}
\caption{Deuteron vector and tensor polarization observables at the energy $E_d=270$ MeV in lab.
The solid and dashed curves are calculations with the DWF derived in the Bonn-CD model and the
DBM respectively. The experimental data are taken from Ref. \cite{RIKEN}.\label{RIKEN1}}
\end{center}
\end{figure}
\begin{figure}[h]
\begin{center}
\includegraphics[width=17cm,keepaspectratio]{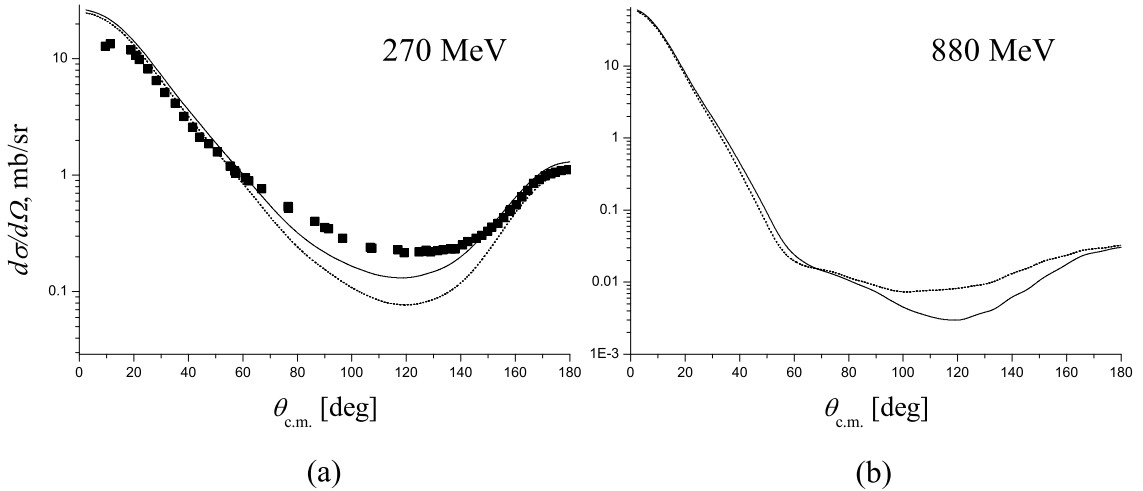}
\caption{The differential cross section at energies $E_d=270$ and $880$ MeV. The definition of
the curves is the same as in Fig. \ref{RIKEN1}. \label{Cross_Sec}}
\end{center}
\end{figure}
\begin{figure}[h]
\begin{center}
\includegraphics[width=17cm,keepaspectratio]{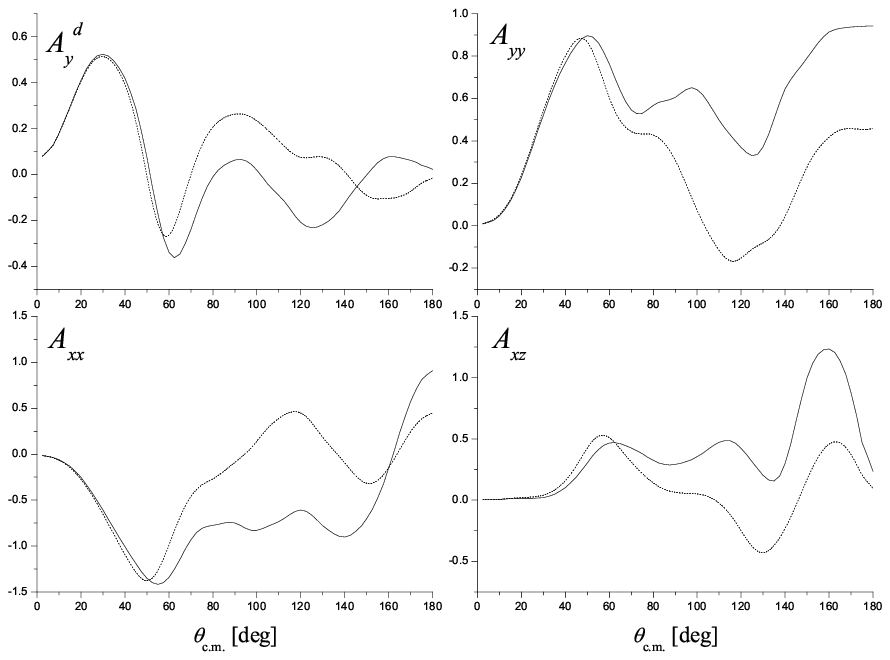}
\caption{The same as in Fig. \ref{RIKEN1}, but for $E_d=880$ MeV.\label{JINR1}}
\end{center}
\end{figure}
 Although the $NN$ T-matrices in the $Nd$ amplitude are taken on-shell,
they still depend on the off-shell momenta, particularly on the internal nucleon momentum in the
deuteron. This dependency is hidden in a value of the effective on-shell energy and $NN$
scattering angle. The integration on the internal momentum means that the knowledge of the
T-matrix at a large energy interval is required. The $NN$ T-matrices are calculated using the
recent partial wave analysis SP07 \cite{Arndt} which extends to 3 GeV for $pp$ scattering and
1.3 GeV for $np$ scattering. All partial waves up to the total angular momentum $J_{NN}=7$ are
taken.

In Figs. \ref{RIKEN1} and \ref{Cross_Sec}(a) the calculations of the observables at the deuteron
energy $E_d=270$ MeV are shown for the two different deuteron wave functions. The experimental
data are taken from Ref. \cite{RIKEN}. The convergent results in the sum (\ref{sum}) are
obtained at $J=25/2$. The full curve is a calculation with the Bonn-CD DWF, and the dashed curve
is a calculation with the DWF from the dressed-bag model. As one can see, the two calculations
do not differ significantly from each other. The difference between them is of the same order as
a disagreement with the experimental data and is seemed to be caused by approximate treatment of
the off-shell effects in the $NN$ T-matrix. Whereas for the Bonn-CD calculation the assumption
of locality of the $NN$ potential and amplitude may be a good approximation, the $NN$ potential
in the DBM is highly nonlocal and energy dependent, thus some off-shell effects from the
$T$-matrix of the DBM may cancellate the effects from the nodal behavior of the DWF. As for the
differential cross-section, the calculations practically coincide with that derived from a
solution of the Faddeev equations without a 3NF. The lack of the cross-section at intermediate
scattering angles is a common feature of such a calculations. Thus the higher rescattering terms
in the $Nd$ optical potential and dynamical off-shell effects are not significant for the cross
section at this energy. The deuteron polarization observables are also in a good agreement with
the experiment. It should be noted here, that even the Faddeev calculations that include a 3NF
are only partially successful in a description of the polarization data.

The convergence of the calculations for forthcoming experimental data at $E_d=880$ MeV
\cite{JINR} is achieved at the total angular momentum $J=39/2$. This is a quite large value,
hence the Faddeev calculations are very difficult to solve at higher energies in the partial
wave basis. As can be seen from Fig. \ref{JINR1}, the differences between the curves derived
from the two DWFs are remarkable. Only for small scattering angles $\theta \leq 50^\circ$, where
the transferred momentum is not large, the two calculations give almost the same results. It is
interesting to note, that at very backward scattering the differential cross-section is mostly
insensitive to the kind of a DFW and the maximum difference becomes apparent in the
cross-section minimum region (see Fig. \ref{Cross_Sec}(b)). However, to make quantitative
calculations at these energies the 3NF contribution due to the excitation of the $\Delta$-isobar
must be taken into account. This contribution is the most likely mechanism to render the
cross-section fall-off at backward angles \cite{Uzikov2}. Furthermore in the dressed-bag model,
$3.6\%$ to the DWF contributes from the $6q$-bag which at this scattering energies can take a
large transferred momentum. So, the 3NF, originated from the scattering of a nucleon on this
quark bag, can also provide a significant contribution at backward scattering angles. Anyway,
such a 3NF provides a large amount of the $^3\rm{H}$ and $^3\rm{He}$ bound energies \cite{Kuk4}.

\section{Conclusions}
A calculation of the deuteron polarization observables $A^d_y$, $A_{yy}$, $A_{xx}$ and $A_{xz}$
and the differential cross section in an optical potential formalism for elastic
nucleon-deuteron scattering at incident deuteron energies $E_d=270$ and $880$ MeV was presented.
Under the investigation was the calculations with two different deuteron wave functions derived
from the Bonn-CD $NN$-potential model and the QCD-motivated dressed bag model. For the $NN$
input, the model independent approach was employed, in which the nucleon-nucleon T-matrix was
taken to be on-shell in a way that some kinematical off-shell effects were incorporated in the
definitions of the effective $NN$ scattering angle and energy. It was found that the
differential cross section is not affected by the higher rescattering contributions and the
off-shell effects in the $NN$ amplitude have a minor influence on this observable. At the energy
$E_d=880$ MeV, large differences in the observables calculated from the two DWFs were observed
in the cross-section minimum region both for the analyzing powers and the differential cross
section. However at the backward scattering angles, the differential cross section was shown to
be mostly insensitive to the short-range deuteron structure.

The author thanks Dr. V.P. Ladygin for helpful discussions and Prof. V.I. Kukulin for providing
a code with the parameterized deuteron wave function of the dressed bag model. This work is
partly supported by the Russian Foundation for Basic Research, grant 07-02-00102a.

\end{document}